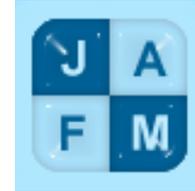

# On the Determination of the Yield Surface within the Flow of Yield Stress Fluids using Computational Fluid Dynamics

N. Schaer[1,2,3†], J. Vazquez[2,3], M. Dufresne[1], G. Isenmann[2,3] and J. Wertel[1]

[1] *3D Eau, Strasbourg, Bas-Rhin, 67000, France*
[2] *Mechanics Department, ICube Laboratory, University of Strasbourg, Strasbourg, Bas-Rhin, 67000, France*
[3] *Department of Fluid Mechanics, National School for Water and Environmental Engineering of Strasbourg (ENGEES), Strasbourg, Bas-Rhin, 67000, France*

†*Corresponding Author Email: nicolas.schaer@3deau.fr*

## ABSTRACT

A part of non-Newtonian fluids are yield stress fluids. They require a minimum stress to flow. Below this minimum value, yield stress fluids remain solid. To date, 1D and 2D numerical models have been used predominantly to study free surface flows. However, some phenomena have three-dimensional behaviour such as the appearance of the limit between the liquid regime and the solid regime. Here the aim is to use a Computational Fluid Dynamics (CFD) to reproduce the properties of the free surface flow of yield stress fluids in an open channel. Modelling the behaviour of the yield stress fluid is also expected. The numerical study is driven with the software OpenFOAM. Numerical outcomes are compared with experimental results from model experiment and theorical predictions based on the rheological constitutive law. The 3D model is validated by evaluating its capacity to reproduce reliably flow patterns. The depth, the local velocity and the stress are quantified for different numerical configurations (grid level, rheological parameters). Then numerical results are used to detect the presence of rigid and sheared zones within the flow.

## NOMENCLATURE

| | | | |
|---|---|---|---|
| $g$ | gravitational acceleration | $\alpha$ | volume fraction |
| $h$ | height | $\dot{\gamma}$ | strain rate |
| $h^+$ | dimensionless height | $\eta$ | viscosity |
| $K$ | fluid consistency | $\eta_0$ | creeping viscosity |
| $n$ | flow index | $\eta_T$ | transition viscosity |
| $p$ | pressure | $\eta^+$ | dimensionless viscosity |
| $q$ | flow rate | $\eta_0^+$ | dimensionless creeping viscosity |
| $t$ | time | $\theta$ | slope |
| $u$ | local velocity | $\rho$ | density |
| $u_r$ | compression velocity | $\tau$ | shear stress |
| $U$ | mean velocity | $\tau_c$ | yield stress |
| $u^+$ | dimensionless velocity | $\tau_w$ | wall shear stress |
| $x$ | distance | $\tau^+$ | dimensionless yield stress |
| $x^+$ | dimensionless distance | | |

## 1. INTRODUCTION

Yield stress fluids occur both naturally (debris flow, avalanches, blood, etc.) and in many industrial fields (petroleum, paints, cosmetics, creams, food, etc.). Their behaviour is characterized by a nonlinear relationship between the stress $\tau$ and the strain rate $\dot{\gamma}$. Their behaviour changes depending on the stress they undergo. When the stress is below the yield stress $\tau_c$, they react as rigid solids. As the stress is higher than the yield value, the material behaves as a liquid and undergoes volume deformations. The behaviour of yield stress fluids is difficult to predict and they are still the subject of research (e.g. Mendes *et al.* 2015; Mendes *et al.* 2017; Mohammadzadeh *et al.* 2016).

The computation of yield stress fluid is a real challenge due to the difficulty to represent the material behaviour in numerical codes. Indeed, the transition from the solid regime to the liquid regime leads to a discontinuity which is complex to describe numerically. To work around this simulation issue, two calculation approaches have been developed (Balmforth *et al.* 2014) such as the Augmented Lagrangian methods (AL method) and the classical formulation with regularized models of the constitutive law.

The AL method involves a variational formulation of the Navier-Stokes equation (Duvaut and Lions 1972; Bristeau and Glowinski 1974; Glowinski *et al.* 1981, Fortin and Glowinski 1983; Glowinski and Le Tallec 1989). The solution of the flow is found thanks to an optimization algorithm. Calculations are difficult to implement because of the nondifferentiability of the dissipation-rate functional at the yield surface. Thus, a related saddle-point problem (Chiang 1984) is introduced in the computational technique to solve this difficulty. The solution is found by iterative approach (Uzawa 1958; Bresch *et al.* 2010; Vola *et al.* 2003; Vola *et al.* 2004). Therefore the discontinuity of the rheological law can be taken into account in calculations.

The regularized models (Bercovier and Engelman 1980; Tanner and Milthrope 1983; Papanastasiou 1987) are another method used to compute yield stress fluids. The main assumption is to consider that the effective viscosity becomes infinite at the yield surface and within a plug. Below the yield stress, the material is represented by a large, but finite viscosity. A large number of experimental configurations have been handled with regularized models (O'Donovan and Tanner 1984; Keentok *et al.* 1985; Ellwood *et al.* 1990; Abdali *et al.* 1992; Mitsoulis *et al.* 1993; Tsamopoulos *et al.* 1996; Blackery and Mitsoulis 1997; Papanastasiou and Boudouvis 1997).

Currently, the AL and the regularized approaches have been developed for finite-element and finite-volume methods. The AL methods are complex to implement but they provide the full stress field in the flow. Nevertheless the method requires high computational power to get a simulation time which is not too long (Roquet and Saramito 2003; Glowinski and Wachs 2011). Regularized models make calculations easier because they ensure a fast convergence (Frigaard and Nouar 2005). A limit to their application is the choice of the regularization parameter which is often set arbitrarily. The CFD solution is difficult to get as the regularization law is close to the real model of a yield stress fluid. Moreover, this approach does not always guarantee to converge to the correct stress field but some works have shown the ability of such regularized models to get yield surfaces for few nontrivial flows (Burgos *et al.* 1999; Alexandrou *et al.* 2001; Nirmalkar *et al.* 2013; Jeong 1013; Thakur *et al.* 2016).

This work presents the application of a regularized model to simulate a free surface flow of a yield stress fluid. The objective is to assess the ability of the code to identify the position of the yield surfaces within the flow. We also discuss the computational limits of the code due to this approach. The present paper is organized as follows. Section 2 presents the case study: an open rectangular channel. Then, Section 3 introduces the numerical model and the dimensionless formulation we used. Finally, the results are discussed in Section 4.

## 2. CASE STUDY

We study the free surface flow of a yield stress fluid in a rectangular open flow channel, which has been widely investigated in the literature (Coussot 1994; Balmforth *et al.* 1999; Balmforth *et al.* 2006). Figure 1 introduces the main geometrical parameters of the straight channel. The slope of the flume is: $\theta = 4.9°$. The material is Carbopol which is a typical non-Newtonian material used for experiments (Cochard 2007; Rudert and Schwarze 2009; Rentschler 2010; López Carranza 2012; Maßmeyer 2013).

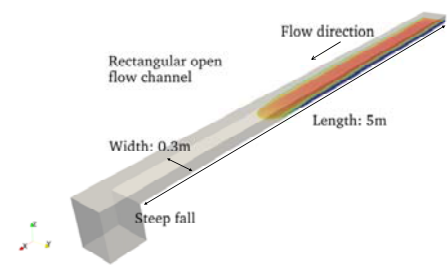

**Fig. 1. Representation of the flow geometry.**

## 3. MATERIALS AND METHODS

### 3.1. Numerical Model

#### 3.1.1. Volume-of-Fluid Approach

The Volume-of-Fluid (VOF) method (Hirt and Nichols 1981) is implemented for the 3D model with the software OpenFOAM©. The time-dependent equations for mass (see Eq. (1)) and momentum (see Eq. (2)), commonly used in fluid mechanics (Versteeg and Malalasekera 2007), are solved with the solver interFoam to represent the free surface of the incompressible, isothermal and immiscible two-phase flow between Carbopol and air.

$$\nabla . U = 0 \qquad (1)$$

$$\rho \frac{\partial U}{\partial t} + \rho \nabla . (UU) = -\nabla p + \rho g + \rho \nabla . (\eta \nabla U) \qquad (2)$$

where $\rho$ (kg/m³) is the density, $U$ (m/s) is the velocity vector, $t$ (s) is the time, $p$ (Pa) is the pressure, $g$ (m/s²) is the gravitational acceleration and $\eta$ (m²/s) is the kinematic viscosity. The VOF method is based on the interface capturing approach. The code involves the volume fraction as a marker to identify the portion of the calculation cell occupied by the fluid. When $\alpha = 0$, the calculation cell is occupied by air. If $\alpha$ is between 0 and 1, it corresponds to the fluid-air interface. Then, if $\alpha = 1$, the calculation cell is occupied by fluid. An advection function allows the transport of $\alpha$. The conservation of volume fraction is essential in particular if the fluid has a high density. In this case, a small error in the volume fraction leads to a significant error on the fluids' properties, and modifies the interface position. As the advection equation cannot satisfy this condition (Rusche 2002), an additional term, called the artificial compression, is introduced in the transport equation of $\alpha$ (Berberović *et al.* 2009; Kissling *et al.* 2010) as shown in Eq. (3):

$$\frac{\partial \alpha}{\partial t} + \nabla \cdot (\alpha U) + \nabla \cdot [u_r \alpha (1 - \alpha)] = 0 \quad (3)$$

where $\alpha$ is the volume fraction, $t$ (s) is the time, $U$ (m/s) is the velocity vector, and $u_r$ (m/s) is the relative velocity perpendicular to the interface between the two phases. The compression velocity is limited to the region close to the interface by the term $\alpha(1-\alpha)$. In the simulations we carried out, the time step is automatically adjusted for the time discretization in order to maintain the Courant number below 1. To ensure the converge of the model to a laminar steady state, the simulations are performed during 60 s.

### 3.1.2. Computational Grids

As Eulerian technique, the VOF method requires a computational grid. Three-dimensional hexahedron grids (see Table 1) were employed to decompose the domain into small finite volume elements.

**Table 1 Computational grids used for the 3D model.**

| Grid level | | Coarse | Medium | Fine |
|---|---|---|---|---|
| Cell size (mm) | $x$-axis | 25.0 | 12.5 | 6.25 |
| | $y$-axis | | | |
| | $z$-axis | | | |
| Number of cells | | 77,000 | 665,000 | 1.15 million |

### 3.1.3. Rheology

The Herschel-Bulkley model describes the rheology of Carbopol (Piau 2007). The strain rate $\dot{\gamma}$ experienced by the material is related to the stress $\tau$ in a non-linear way (Herschel and Bulkley 1926) as Eq. (4) and Eq. (5) show:

$$\tau = \tau_c + K.\dot{\gamma}^n \; for \; \tau > \tau_c \quad (4)$$

$$\dot{\gamma} = 0 \qquad for \; \tau \leq \tau_c \quad (5)$$

where $\tau_c$ is the yield stress (Pa), $K$ is the fluid consistency (Pa.s$^n$) and $n$ is the flow index. In the present study, Carbopol is described with the following parameters: $\tau_c$ =7.5 Pa, $K$ = 4.6 Pa.s$^n$, and $n$ = 0.39. These values were determined from rheological measurements (Debiane 2000). The density $\rho$ is 1,000 kg/m$^3$. In CFD, it is not possible to employ such discontinuous material behaviour. The code can solve the equations of fluid mechanics but not the ones for pure solid mechanics (Greenshields 2015). It cannot fully considerer the solid regime when the stress is under the yield stress. To overcome this simulation issue, we used the bi-viscosity regularized model (Tanner and Milthrope 1983) implemented in the code as follows:

$$\eta = min(\eta_0, \frac{\tau_c}{\dot{\gamma}} + k.\dot{\gamma}^{n-1}) \quad (6)$$

where $\eta$ is the kinematic viscosity, $\eta_0$ is the creeping viscosity and $\dot{\gamma}$ is strain rate. $\tau_c$, $K$, and $n$ are the conventional parameters of the Herschel-Bulkley model. The regularization considers the value of the viscosity is very large ($\eta_0$), yet finite, when the fluid behaves as a solid ($\tau < \tau_c$). Beyond the yield stress ($\tau > \tau_c$), the viscosity in the computational cell is determined by a power law as described in Eq. (6). The literature does not provide any typical value for $\eta_0$. As an example, for a Bingham fluid, $\eta_0$ should be about 1,000 times greater than the viscosity in the liquid regime (O'Donovan and Tanner 1984). The definition of $\eta_0$ is to some extent arbitrary and not based on exact measurements of the yield stress (Rudert and Schwarze 2009). Thus we investigate the effect of the creeping viscosity $\eta_0$ on the numerical results.

### 3.1.4. Boundary Conditions and Initial Values

In the present case, the discharge is constant ($q = 3$ l/s). In the model, a velocity condition is set upstream the channel as a Dirichlet condition type. At the bottom of the fall, the output of the model is considered to be at atmospheric pressure (the same condition is set for the top of the model). A Dirichlet type boundary condition was given for $\alpha$ except for the walls. Note that all wall boundaries in the flow region obey the no-slip condition. More details about boundary conditions are provided in Table 2.

**Table 2 Boundary conditions for inlet, outlet, top and walls.**

| Boundary | Field | Type | Value |
|---|---|---|---|
| Inlet | $\alpha$ | Dirichlet | $\alpha = 1$ |
| | $p$ | Neumann | $\underline{n}.\nabla p = 0$ |
| | $U$ | Dirichlet | $q_{in} = 3 \; l/s$ |
| Outlet | $\alpha$ | Dirichlet | $\alpha = 0$ |
| | $p$ | Dirichlet | $p_{out} = 0$ |
| | $U$ | Neumann | $\underline{n}.\nabla U = 0$ |
| Top | $\alpha$ | Dirichlet | $\alpha = 0$ |
| | $p$ | Dirichlet | $p_{out} = 0$ |
| | $U$ | Neumann | $\underline{n}.\nabla U = 0$ |
| Wall | $\alpha$ | Neumann | $\underline{n}.\nabla \alpha = 0$ |
| | $p$ | Neumann | $\underline{n}.\nabla p = 0$ |
| | $U$ | Dirichlet | $\underline{U} = 0$ |

### 3.1.5. Application of the Numerical Model

First, a grid sensitivity study is carried out to quantify the uncertainties due to the computational grids. Second, we assessed the sensitivity of the numerical results to the rheological parameters. A test with a variation of the channel slope is performed to validate the set-up determined from the sensitivity studies. The CFD results are compared with published experimental data (Debiane 2000) and analytical models (Coussot 1994; Piau 1996; Burger *et al.* 2010). These first three steps validate the numerical model with the configuration with the lowest deviations from the experimental data. Then, we investigate the position of yield surfaces within the flow by evaluating the viscosity, velocity and the stress fields.

### 3.2. Dimensionless Formulation

In this paper, the following dimensionless variables will be used:

$$u^+ = \frac{u}{u^*}, \tag{7}$$

$$h^+ = \frac{h}{h_0}, \tag{8}$$

$$x^+ = \frac{x}{L}, \tag{9}$$

$$\tau^+ = \frac{\tau}{\tau_c}, \tag{10}$$

and

$$\eta^+ = \frac{\eta}{\eta^*} = \frac{\eta}{u^* h_0}, \tag{11}$$

with

$$u^* = \left(\frac{\tau_c}{\rho}\right)^{\frac{1}{2}}, \tag{12}$$

and

$$h_0 = \frac{\tau_c}{\rho g \sin \theta}, \tag{13}$$

the velocity and height scales, respectively. In fluid mechanics, $u^*$ is called the shear velocity. It is not an actual velocity but it is a quantity involving the boundary shear stress that conveniently has the dimensions of a velocity. So, $u^+$ is the ratio between the local velocity $u$ (m/s) and $u^*$. $h_0$ corresponds to the minimum height required to generate a flow in an inclined channel (Coussot 1994). Indeed, $h^+$ is the ratio between the height of the flow $h$ (m) and $h_0$. The dimensionless variable $x^+$ is defined by the distance $x$ (m) run by the flow and the total length of the channel $L$ (here $L$ = 5 m). The shear stress $\tau$ (Pa) is introduced with $\tau^+$ and the kinematic viscosity $\eta$ with $\eta^+$.

### 4. RESULTS AND DISCUSSION

#### 4.1. Computations for Model Validation

##### 4.1.1. Grid Sensitivity Study

The quality of the mesh is assessed according to the regularity of the finite volume elements. This is used to minimize the cells' distortion (skewness). Computational approaches such as the VOF method lead to uncertainties due to the numerical model used. This error comes from the discretization of the computational domain in small finite size elements. Thus we carried out a grid sensitivity study to quantify the numerical uncertainties estimated by the Grid Convergence Index (GCI). This method is proposed by Roache (Roache 1994). The three grids are used independently in simulations. The rheology is fixed with the values described in Section 3.1.3. The value of $\eta_0$ is fixed arbitrarily in the regularization model: $\eta_0^+$ = 2.6e4. The objective is to validate the model for an acceptable level of mesh providing the lowest deviation with experimental data and minimum numerical uncertainties. Figure 2 presents the CFD results in a dimensionless form for the velocity. The profiles are plotted in the vertical direction. The results for the coarse grid are the worst. Both the velocity field and the depth are too far away from experimental data to consider the coarse mesh usable to the rest of the study.

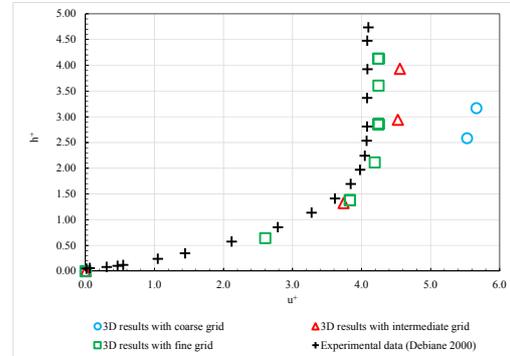

**Fig. 2. Grid sensitivity study: velocity profiles in the vertical direction.**

This large difference can be explained by the few number of calculation cells on the vertical axis (only 2 cells). The intermediate mesh and the fine mesh are more reliable compare to experimental data even if they tend to overestimate the maximum velocity in the upper part of the flow. By reducing the cell size, the gap between the maximum velocity of numerical and experimental results is reduced from 11 % (intermediate mesh) to 4 % (fine mesh). As the results for fine mesh show, the material remains rigid near the free surface ($h^+$ > 2.50). At the bottom of the channel, the flow is quickly sheared and deformed when the velocity gradient increases. In the plug region ($h^+$ > 2.50; $u^+$ > 4.0), the velocity is constant and the shear stress is theoretically below the yield stress $\tau_c$ with a value close to zero but nonzero (Coussot, 1994; Balmforth and Craster 1999). At this stage of the study, we note that the behaviour of the Carbopol, as a yield stress fluid, is well reproduced by the model.

The backwater curves are represented in Fig. 3. In the upstream part ($x^+$ < 0.80), the flow regime tends to be uniform. The three meshes present some deviations with experimental data ($h^+$ = 4.73) in terms of depth: $h^+$ = 3.15 for the coarse mesh, the intermediate grid provides a value of normal depth of $h^+$ = 3.88. Once again, the number of cells of the coarse grid degrades the flow representation compared to the other meshes. The lowest gap is observed for the fine mesh ($h^+$ = 4.12). Near the fall downstream ($x^+$ > 0.80), the flow is accelerated. The depth decreases gradually and reaches the critical depth in a cross section close to the fall. As Fig. 3 shows, the shape of the free surface for the fine mesh is the closest to the experimental free surface. The results are also compared with an analytical model (Piau, 1996). This model is an extension of the Saint-Venant expression. It takes into account the influence of normal stress on the *y*-axis due to yield stress. The intermediate mesh results are quite similar to the depth computed by the analytical model in the zone where the flow is uniform. But, in the area of the fall, the analytical model tends to the experimental data and it is the fine mesh that shows lowest gaps.

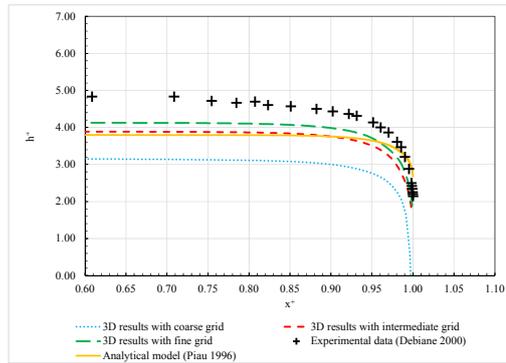

**Fig. 3. Grid Sensitivity Study: Backwater Curves.**

To complete the assessment of the model, the CFD normal depth in the part where the flow is uniform is faced to two other laws, proposed by Coussot (Coussot 1994) and Burger (Burger *et al.* 2010). They have been established for cases of yield stress flows in rectangular flumes. Table 3 present the comparison. The deviations between CFD results and experimental data are also provided. A deviation of about 15 % is obtained with the fine mesh as Fig. 3 shows previously. But the normal heights calculated by the laws are closer to the results of the intermediate mesh.

**Table 3 Grid sensitivity study: comparison of CFD normal depths with analytical models.**

| Mesh level | Coarse | Inter | Fine |
|---|---|---|---|
| CFD | 3.15 | 3.88 | 4.12 |
| Coussot model | | 3.59 | |
| Burger model | | 3.80 | |
| Exp. value | | 4.73 | |
| Deviation CFD /exp. | 50.2 % | 21.9 % | 14.8 % |

We assume at this stage that the differences can be explained by the numerical uncertainties associated with mesh size. Indeed, the GCI is evaluated to quantify these uncertainties (see Table 4). The accuracy of the model for the depth is 5 % for the fine mesh and 19 % for the intermediate mesh. Thus the height with fine mesh is assumed to be between 3.91 and 4.33.

**Table 4 GCI for the maximum velocity and the depth in numerical simulations.**

| Mesh | Maximum $u^+$ | GCI for $u^+$ | $h^+$ | GCI for $h^+$ |
|---|---|---|---|---|
| Coarse | 5.66 | - | 3.15 | - |
| Inter. | 4.56 | 24 % | 3.88 | 19 % |
| Fine | 4.25 | 7 % | 4.12 | 5 % |

As a conclusion of the grid sensitivity study, the CFD results show significant differences with experimental data. For a fixed rheology, we observed the deviation with the experiment is reduced as the computation cell size decreases. The intermediate mesh, with 3 cells on the vertical axis, seems to be more reliable to analytical models for the uniform regime: 9 % gap with the Coussot model and 3 % gap with the Burger law. But the quality of the grid leads to high numerical uncertainties (GCI for $u^+$ is 24 %; GCI for $h^+$ is 19 %) which are reduced with the fine mesh. The last grid we tested (6 cells on the vertical axis) is more precise to reproduce the behaviour of the yield stress fluid both for the velocity vertical distribution and the shape of the free surface when the regime is gradually varied near the fall. Besides the numerical depth is fairly close to the one estimated by the Coussot model (difference of 13 %) and the Burger law (difference of 8 %). Therefore the fine mesh is retained for the rest of the study even if the intermediate grid hence a faster computation time. The fine mesh is used for the rheology study introduced in the next section.

### 4.1.2. Rheology Sensitivity Study

Although the gap between CFD results and experimental data is reduced with the fine mesh, the deviation remain significant for the depth (see Fig. 3). For the grid study, we used the rheology that was measured before the experiment in the channel (Debiane 2000). The typical errors associated with rheometrical measurements lead to uncertainties on the order of 30 %. Difference appear on rheological parameters when they are re-evaluated on the base of the flow characteristics observed during the experiments. The yield stress increases by 14 % and the consistency decreases by 4 % (Debiane 2000). Only the flow index remains constant. Thereby we assessed the sensitivity of the CFD results to the changes of the Herschel-Bulkley parameters. The objective is to validate the model for the rheological set-up including the lowest deviations with experimental data. The mesh is fixed with the fine grid to study the variations of the rheological parameters independently. We arbitrarily varied each parameter by +30 % and -30 % (relative to experimental values). For all simulations the creeping viscosity $\eta_0$ is arbitrarily fixed in the regularization law ($\eta_0^+$ = 2.6e4). Table 5 summarizes the configurations we tested.

**Table 5 Rheological parameters for the sensitivity study.**

| Configuration | $\tau+$ | K | n |
|---|---|---|---|
| Exp. rheology | 1 | 4.60 | 0.39 |
| + 30 % on $\tau_C$ | 1.3 | 4.60 | 0.39 |
| - 30 % on $\tau_C$ | 0.7 | 4.60 | 0.39 |
| + 30 % on $K$ | 1 | 5.98 | 0.39 |
| - 30 % on $K$ | 1 | 3.22 | 0.39 |
| + 30 % on $n$ | 1 | 4.60 | 0.507 |
| - 30 % on $n$ | 1 | 4.60 | 0.273 |

Similar to mesh study, we plotted backwater curves (see Figs. 4, 5, 6) and estimated the normal depths (see Table 6).

We note the results are sensitive to the rheology as they differ from those obtained with experimental rheology (see Sec. 4.1.1). Figure 7 sums up the sensitivity of the results by introducing the depth variation as function of the gap with the experimental data. Changing the value of the yield stress leads to the lowest variations in terms of depth compared to the CFD with experimental rheology. The depth increases by about 2 % when $\tau_c$ increases by 30 % and decreases by about 12 % when $\tau_c$ is lower by 30 % (see. Table 6 and Fig. 7). At best, the model underestimates the normal depth by 13 % compared to the experimental value. The shape of the numerical backwater curve remains different from the experimental free surface regardless the variation of $\tau_c$. The results for the consistency $K$ have greater depth variations than for $\tau_c$. The depth increases by 9 % when the value of $K$ is 5.98. On the contrary reducing $K$ by 30 % leads to a decrease of about 21 % in normal depth. Rising the consistency also causes smaller deviations with the experiments (the model underestimates the normal depth by 5.3 %) and improves the shape of the numerical free surface. The tests carried out on the flow index $n$ lead to the greatest depth variations in comparison with CFD using experimental rheology.

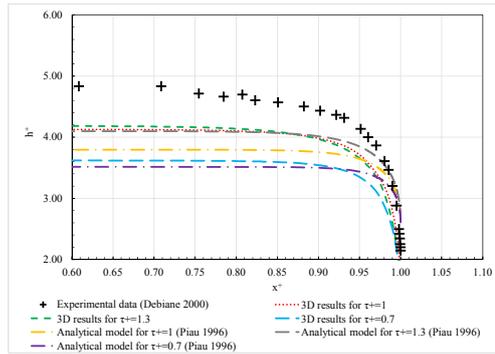

**Fig. 4. Yield stress sensitivity study: backwater curves.**

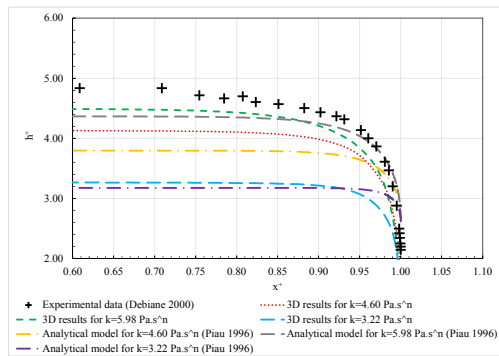

**Fig. 5. Consistency sensitivity study: backwater curves.**

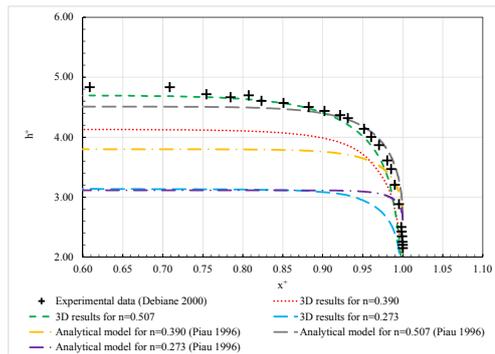

**Fig. 6. Flow index sensitivity study: backwater curves.**

**Table 6 Rheology sensitivity study: comparison of CFD normal depths with analytical models.**

| Rheology | $\tau^+ = 1$ | $\tau^+ = 1.3$ | $\tau^+ = 0.7$ |
|---|---|---|---|
| CFD | 4.12 | 4.19 | 3.62 |
| Coussot model | 3.59 | 3.82 | 4.19 |
| Burger model | 3.80 | 4.10 | 3.52 |
| Exp. value | 4.73 | | |
| Deviation CFD /exp. | 14.8 % | 12.9 % | 30.7 % |
| Rheology | $K = 4.60$ | $K = 5.98$ | $K = 3.22$ |
| CFD | 4.12 | 4.49 | 3.27 |
| Coussot model | 3.59 | 3.91 | 3.33 |
| Burger model | 3.80 | 4.37 | 3.17 |
| Exp. value | 4.73 | | |
| Deviation CFD /exp. | 14.8 % | 5.3 % | 44.6 % |
| Rheology | $n = 0.390$ | $n = 0.507$ | $n = 0.273$ |
| CFD | 4.12 | 4.70 | 3.14 |
| Coussot model | 3.59 | 3.89 | 3.37 |
| Burger model | 3.80 | 4.51 | 3.11 |
| Exp. value | 4.73 | | |
| Deviation CFD /exp. | 14.8 % | 0.6 % | 50.6 % |

A depth divergence of 14.1 % is noticed (Fig. 7) when $n$ is set with +30 % on its value. On the other hand, reducing the parameter by 30 % provides a drop of more than 24 % in depth. So, the numerical results are highly sensitive to the variations of the flow index. Figure 6 shows that a value of $n = 0.507$ produces a backwater curve close to the experimental free surface. The analytical curve with the same value of flow index has small deviations with the CFD, especially near the fall where the material is accelerated. That suggests the model is pretty accurate in this part of the flow. It is also noted that the deviation for the normal depth is less than 1 % with the measured value, which is very satisfying.

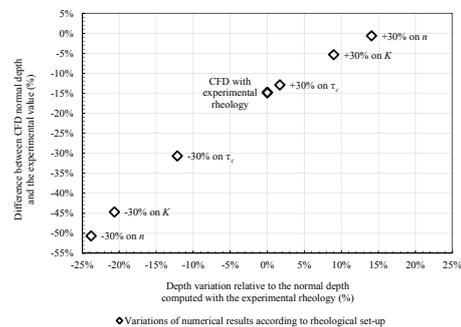

**Fig. 7. Rheology sensitivity study: variations of the CFD results.**

As a conclusion of the rheology sensitivity study, the gap with the experimental data can be strongly reduced by increasing of 30 % the flow index. Thus we decided to retain this configuration for the rest of present study.

### 4.1.3. Slope Sensitivity Study and Validation of the Model

To validate the model with the fine mesh and the rheology we set (+ 30 % on the flow index), we carried out a numerical test by reducing the slope of the channel from 4.9° to 2.4°. The results we obtained are presented in Figs. 8, 9 and Table 7. They are very close to the experimental data, especially in terms of depth. The difference is 10 % for the normal depth. The backwater curve has a similar shape to the real free surface particularly in the zone near the fall. On the velocity profile an overestimation of the maximum velocity of 9 % is noted.

The accuracy of the results and their gap to experimental data (maximum gap of 10 %) validate the conclusions of the sensitivity analyses introduced previously. The complementary choice of a fine mesh with a rheology increased by 30 % on the flow index leads to a good representation of the behaviour of the material in the channel. The 3D model is validated for this set-up and used for the detection of the yield surfaces within the flow.

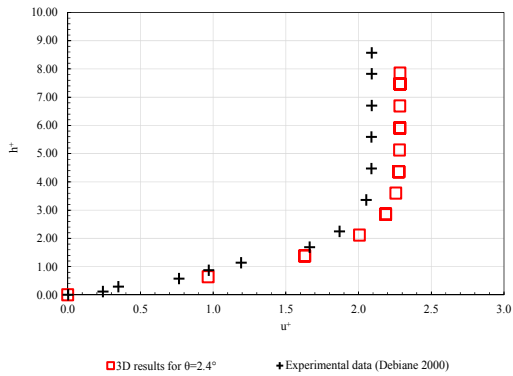

**Fig. 8. Slope sensitivity study: velocity profiles in the vertical direction.**

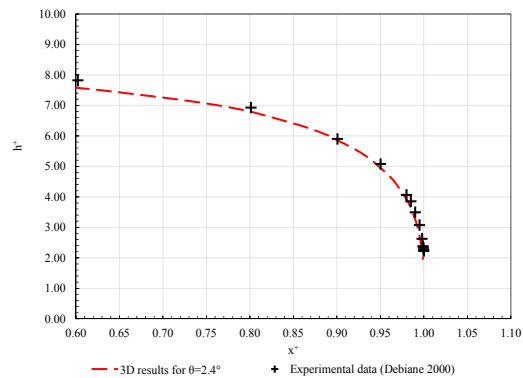

**Fig. 9. Slope sensitivity study: backwater curves.**

**Table 7 Slope sensitivity study: comparison of CFD normal depths with analytical models.**

| Channel slope $\theta$ | 2.4° | 4.9° |
|---|---|---|
| CFD | 7.77 | 4.70 |
| Coussot model | 6.57 | 3.89 |
| Burger model | 7.74 | 4.51 |
| Exp. value | 8.56 | 4.73 |
| Deviation CFD /exp. | 10.2 % | 0.6 % |

### 4.2. Detecting the Yield Surface Within the flow

#### 4.2.1. Suggestion and Validation of a Criterion Based on a Transition Viscosity

From the regularization law used in the code (see Sec. 3.1.3) we assumed a transition viscosity $\eta_T$ to point the passage from the rigid regime to the liquid regime of the material. It is defined as follows:

$$\eta_T = K \cdot n \cdot \llbracket (\tau_c / \eta_0) \rrbracket^{(n-1)}. \quad (14)$$

Therefore, according the viscosity profile in the vertical direction for a fixed distance $x^+$ in the centre of the channel ($y = b/2$), we assumed that the transition between the two zones is located for the depth where the viscosity reaches the value of $\eta_T$. By repeating this technique along the longitudinal axis of the channel, the yield interface is identified by plotting a curve that connects all the transition points of each vertical viscosity profile. We tested the sensitivity of the results to the creeping viscosity. Indeed, the transition viscosity changes for each test. Five different values have been evaluated: $\eta_0^+ = 3.0e3$, $\eta_0^+ = 2.6e4$, $\eta_0^+ = 2.6e5$, $\eta_0^+ = 2.6e6$ and $\eta_0^+ = 1.2e7$. The classical parameters are set according the model validation ($\tau^+ = 1$; $K = 4.60$; $n = 0.507$). For a slope $\theta$ of 4.9°, the results are introduced by Fig. 10 for $\eta_0^+ = 3.0e3$, by Fig. 11 for $\eta_0^+ = 2.6e5$ and by Fig. 12 for $\eta_0^+ = 1.2e7$. The red dotted line represents the transition viscosity and separates the plug region which appears in grey in the charts. The black dotted lines and the blue lines are the vertical velocity profiles and the backwater curves respectively.

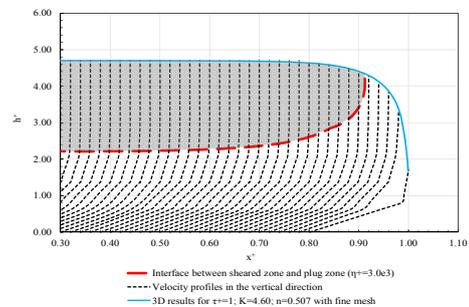

**Fig. 10. Numerical yield surface based on the criterion of the transition viscosity for $\eta_0^+=3.0e3$.**

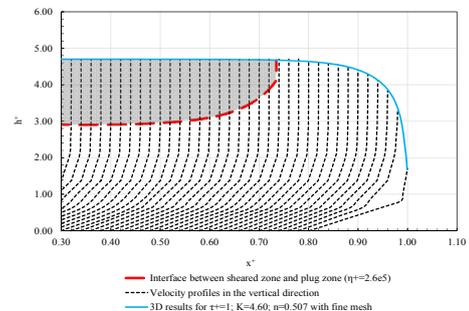

**Fig. 11. Numerical yield surface based on the criterion of the transition viscosity for $\eta_0^+=2.6e5$.**

The yield interface which separates the plug zone and the sheared zone moves according to the creeping viscosity we retained in the regularization law. The plug zone is observed in the upper part of the flow. This is consistent with the characteristics of the flow described in the literature (Coussot 1994; Balmforth and Craster 1999). The plug zone tends to move back upstream the channel and its thickness is reducing when the creeping viscosity increases. We note the shape of the rigid zone is consistent with the theory (Piau 1996). The normal stress decrease from the fall to upstream, up to reaching the point where the yield surface meets the backwater curve. This meeting point is observed at $x^+ = 0.92$ for $\eta_0^+ = 3.0e3$, at $x^+ = 0.74$ for $\eta_0^+ = 2.6e5$ and at $x^+ = 0.54$ for $\eta_0^+ = 1.2e7$. Downstream from this point, the fluid flows because it is sheared and elongated. Upstream from the point, flow depth is equivalent to the depth for a uniform flow. The thickness of the rigid zone increases gradually upstream up to reach the uniform flow properties. At this moment, only shear stress is effective. Normal stress is equal to zero. The flow in the gradually varying region is both sheared and elongated. We also find that the backwater curves and the velocity profiles do not vary regardless the value of the creeping viscosity. Figures 10, 11 and 12 indicate that the variation of the depth in the direction of the flow is identical from one test to another. The velocity profiles are also regular as shown in Fig. 13, which introduces the results for all five tests for a the distance $x^+ = 0.6$. A significant curvature also appears on the velocity profiles near the fall ($x^+ > 0.9$). It means that there is a local acceleration of the flow. In this part, the wall shear stress gradually increases which causes the material to be elongated. This phenomenon is observed in Fig. 14. The variation of the wall shear stress in the direction of the flow is evaluated for $\eta_0^+ = 3.0e3$. The value of the wall shear stress is about $\tau^+ = 4.0$ in the zone where the regime is uniform. Then from $x^+ = 0.95$, it increases to a maximum value of $\tau^+ = 5.6$ at the fall. A comparison with an analytical model (Coussot 1994) is provided to validate the order of magnitude of the CFD results.

At first sight, the detection of the yield surface is conceivable from the criterion based on the transition viscosity $\eta_T$. However, we observe the viscosity field in the computational domain remains rather constant regardless of the $\eta_0$ value we set (see Fig. 15). The viscosity increases when we move from the bottom of the channel to the free surface. This is consistent with the presence of the plug in the upper part of the flow. The viscosity reaches the same maximum value of $\eta^+ = 5.1e3$ except when the regulation parameter is $\eta_0^+ = 3.0e3$. In this case, the viscosity distribution differs from other profiles from a depth of $h^+ = 2.20$ and the maximum value is 2.2e3. In fact the maximum value the viscosity field can reached is entirely dependent of the value of $\eta_0$ we set in the regularization model of the code. Of course, the maximum value depends on the local flow characteristics (velocity, shear) but the regularization parameter $\eta_0$ acts as an arbitrary cut-off point on the viscosity field. This is a limitation to the use of regularized models (Bercovier and Engelman 1980; Tanner and Milthrope 1983; Papanastasiou 1987). Therefore the criterion based on the transition viscosity $\eta_T$ we proposed above is not appropriate since it is fully dependent on the arbitrary value given to $\eta_0$ in the regularization approach.

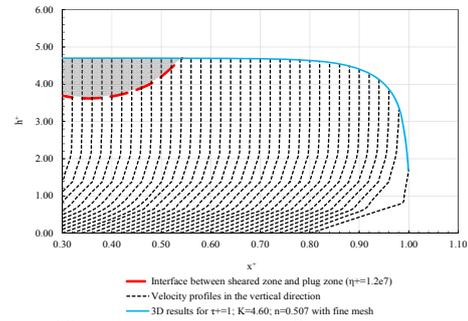

**Fig. 12. Numerical yield surface based on the criterion of the transition viscosity for $\eta_0^+=1.2e7$.**

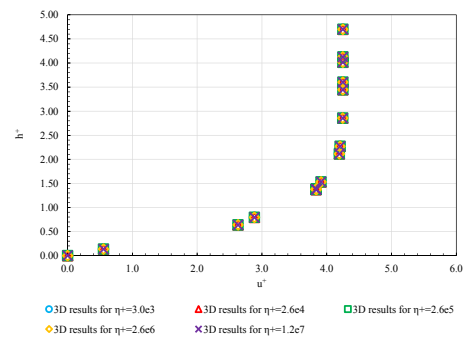

**Fig. 13. Creeping viscosity sensitivity study: velocity profiles in the vertical direction for $x^+ = 0.6$.**

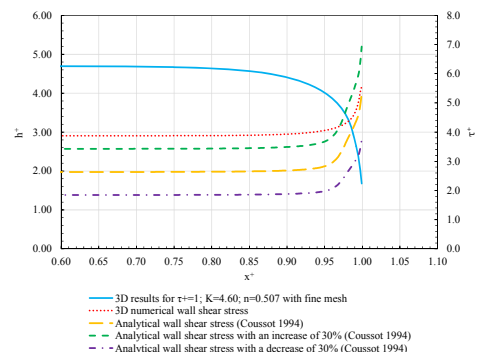

**Fig. 14. Wall shear stress for $\eta_0^+=3.0e3$.**

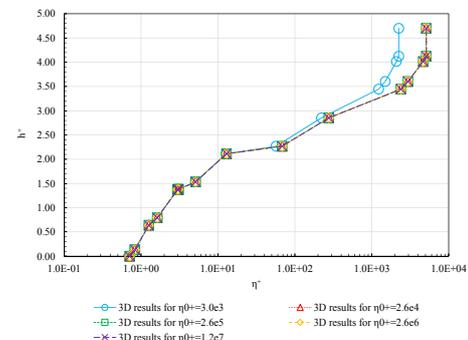

**Fig. 15. Creeping viscosity sensitivity study: viscosity profiles in the vertical direction for $x^+ = 0.6$.**

### 4.2.2. Tracking the Yield Surface from the Velocity Profiles

Due to the unreliability of the transition viscosity criterion, we suggest a second approach based on the velocity profiles. The velocity distribution (see Figs. 2, 8, 13) indicates the two characteristic parts of the yield stress fluid. A pseudo-yield interface can be found by assuming that the plug zone is reached for a depth where the velocity gradient tends to value close to zero. Thus, the transition point must be reached when the velocity magnitude is very close to its maximum intensity. In our case, we assumed that the transition is achieved when the velocity is 99 % of its maximum value. Then it is possible to determine the depth within the flow corresponding to this condition. By repeating this operation for all profiles, a pseudo yield interface appears in the flow as Fig. 16 shows for a value of $\eta_0^+ = 1.2e7$. The position of the interface varies irregularly within the flow according to the distance. The description of the rigid zone is correct because it is near the free surface, but it does not lead to a clear and precise yield interface.

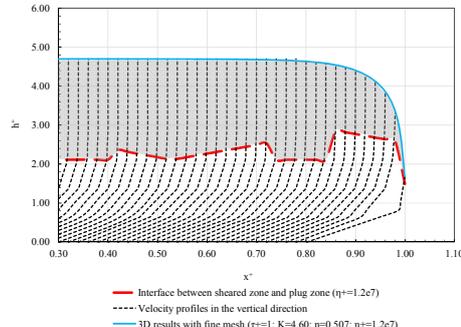

**Fig. 16. Numerical yield surface based on the velocity profiles for $\eta_0^+$=1.2e7.**

The shape of the interface is consistent with the literature (Coussot 1994, Piau 1996) which specifies that the liquid zone to rigid zone transition can be pointed along the vertical velocity distribution only if the flow is steady and uniform. In our case, this condition is met only on the upstream part of the canal. Near the fall, the flow gradually becomes varied and it is no longer possible to use the velocity profiles in the same way.

### 4.2.3. Delimitation of the Yield Surface from the Stress

We assessed a third method to identify the yield surface as it cannot be determined from the viscosity or velocity field. This approach uses the yield stress criterion to point the interface. The stress field is computed at the end of the simulation in the entire domain as follows:

$$|\tau| = \sqrt{\tfrac{1}{2}\{\tau:\tau\}}, \qquad (15)$$

$$|\tau| = \left(\frac{\tau_{xx}^2+\tau_{yy}^2+\tau_{zz}^2+2(\tau_{xy}^2+\tau_{xz}^2+\tau_{yz}^2)}{2}\right)^{0.5}. \qquad (16)$$

We obtained the magnitude of the stress, including the normal stress components ($\tau_{xx}$, $\tau_{yy}$, $\tau_{zz}$) and the shear stress components ($\tau_{xy}$, $\tau_{yz}$, $\tau_{xz}$). Then, all the points taking the value of the yield stress (here $\tau_C =$ 7.5 Pa) along the longitudinal axis are connected to make the yield interface appear. This method has been tested with the five values of the creeping viscosity. The results are introduced in Fig. 17 for $\eta_0^+ = 3.0e3$. The plug zone still appears in the upper part of the flow. Its shape and its size vary when the value of $\eta_0^+$ changes. But we note that above a value of $\eta_0^+$=2.6e4, the plug zone is no longer modified as Fig. 18 indicates.

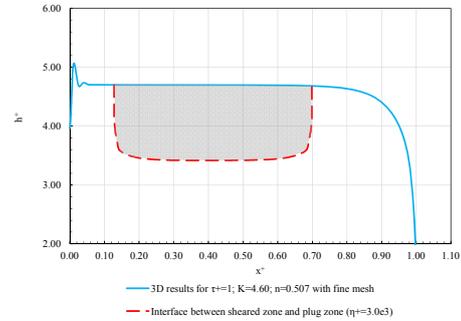

**Fig. 17. Numerical yield surface based on the criterion of the yield stress for $\eta_0^+$=3.0e3.**

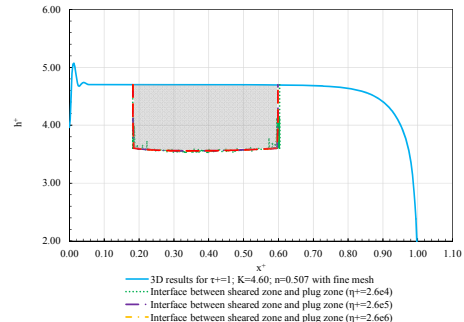

**Fig. 18. Numerical yield surface based on the criterion of the yield stress for $\eta_0^+$=2.6e4, $\eta_0^+$=2.6e5, $\eta_0^+$=2.6e6 and $\eta_0^+$=1.2e7.**

According the results we got, the criterion based on the yield stress is the most appropriate to delimit the yield interface. However, its position is sensitive to the value of $\eta_0$, especially when it is set too low. Therefore we recommend to retain a large value (above 1.0e7) to make the CFD results independent of the regularization parameter in the code.

## 5. CONCLUSION

The free surface flow of a Herschel-Bulkley liquid (Carbopol) into a rectangular channel is investigated numerically with a 3D approach. This type of flow has been widely studied in experimental configurations and it is similar to some natural phenomenon such as the passage of the debris flow in narrow channels or industrial processes (e.g. cast iron flows). The objective of this work was to assess the ability to reproduce the flow with a rheological approach using the regularized bi-viscous model, and to investigate the possibility of detecting liquid and rigid zones using several criteria. In the computations, the flow characteristics have been visualized by tracking the free surface and the yield interface. The rheology of the Carbopol is given by the bi-viscous model which is an approximation of the constitutive equation of a Herschel-Bulkley liquid. All equations are solved with the finite-volume method and the free surface is represented by the Volume-Of-Fluid method. The OpenFOAM software package is employed to integrate the model equations. We note the velocity field and the flow are sensitive to the numerical set-up. Moreover, the velocity profiles in the vertical direction and the backwater curves depend of the quality of the mesh employed. We conclude the position of the yield interface is highly sensitive to the set-up of the regularized model. The patterns of the rigid zone cannot be evaluated according a criterion based on a cut-off of a value of viscosity or by post-processing the velocity profiles. From this wok, it can be concluded a stress criterion is more appropriate to identify the yield interface but it requires a high value of the regularized parameter $\eta_0$ in the code to minimize its influence on the results.

In the future the code development should be focused on the computation of natural flows involving yield interfaces (e.g. debris flow) or on the test of other regularized models. It would be also interesting to investigate the separate effect of Bingham number or Reynolds number. Finally, the method we assessed would be extended to comparisons with experimental data providing real interfaces (Luu et al. 2015; Souza Mendes et al. 2007). Current research is focused on this last point and on the application of the code for debris flow simulation. It will be the subject of future communications (Schaer 2018, to be published).

## ACKNOWLEDGEMENTS

The authors would like to gratefully acknowledge the ANRT, the French national association for research and technology, for funding the 3D Eau Company for this research, and the ICube laboratory of the University of Strasbourg (France) and the National School for Water and Environmental Engineering of Strasbourg (ENGEES) for hosting this work.